\documentclass{IOS-Book-Article}

\usepackage{mathptmx}

\usepackage{graphicx}
\usepackage{subcaption}
\usepackage[normalem]{ulem}
\useunder{\uline}{\ul}{}
\usepackage{adjustbox}
\usepackage{import}
\usepackage[utf8]{inputenc}
\usepackage{csvsimple}
\usepackage{multirow}
\usepackage{tabularx}
\usepackage{flafter}
\usepackage{url}
\usepackage{todonotes}
\usepackage{comment}

\def\hb{\hbox to 10.7 cm{}}

\begin{document}

\pagestyle{headings}
\def\thepage{}

\begin{frontmatter}              

\title{Stop Illegal Comments: A Multi-Task Deep Learning Approach}


\author{\fnms{Ahmed} \snm{Elnaggar}%
\thanks{Corresponding Author: Ahmed Elnaggar, Software Engineering for Business Information Systems, Boltzmannstr. 3, 85748 Garching bei München, Germany; E-mail: ahmed.elnaggar@tum.de.}},
\author[A]{\fnms{Bernhard} \snm{Waltl}},
\author[A]{\fnms{Ingo} \snm{Glaser}},
\author[A]{\fnms{Jörg} \snm{Landthaler}},
\author[A]{\fnms{Elena} \snm{Scepankova}}
and
\author[A]{\fnms{Florian} \snm{Matthes}}

\address[A]{Software Engineering for Business Information Systems, Technische Universität München, Germany}

\begin{abstract}
Deep learning methods are often difficult to apply in the legal domain due to the large amount of labeled data required by deep learning methods. A recent new trend in the deep learning community is the application of multi-task models that enable single deep neural networks to perform more than one task at the same time, for example classification and translation tasks. These powerful novel models are capable of transferring knowledge among different tasks or training sets and therefore could open up the legal domain for many deep learning applications. In this paper, we investigate the transfer learning capabilities of such a multi-task model on a classification task on the publicly available Kaggle toxic comment dataset for classifying illegal comments and we can report promising results.
\end{abstract}

\end{frontmatter}

\section{Introduction}
The analysis of textual documents is an important task in the legal domain~\cite{ashley2017artificial}. A plethora of different use cases exist, that heavily rely on analysis and inspection of documents. The importance is not restricted to legal research tasks, such as the analysis of statutory texts, judgments, or contracts, but also includes reviewing the content of text documents with regard to facts and evidences~\cite{ashley2017artificial}. More and more pieces of circumstantial evidence are discovered using technology, especially when it comes up to inspect huge document corpora. The field of e-Discovery, forensics, legal reasoning and argument mining, and information extraction (IE) is well-studied and established in the field of legal informatics \cite{ashley2010emerging}. Especially the usage during due diligence is highly attractive as it helps to save valuable resources, such as money and time. 

Technology assists human experts during complex discovery tasks to find the “needle in the haystack”. Methods and software tools have been used successfully to detect crime such as organized manipulation, e.g., analysis of a very large collection of e-mails during the VW scandal of manipulated software, or to automatically unveil discrimination in the internet. The latter is especially relevant for social media platforms but also for e-participation initiatives. The past has shown, that anonymity within the internet attracts users for inappropriate, unconstitutional and illegal comments. For example, extremist statements, threats, insults, and so forth. Online platforms are getting more responsible in charge of the content that users create and share with others. The responsibility to delete comments on request but also to proactively delete illegal comments is more and more in the charge of the platform providers, which struggle at this complex and tedious task. This leads to the need of a highly accurate software to perform this task automatically.

This paper contributes to the detection and classification of statements and comments with regard to their sentimental content. The sentiment analysis is restricted to inspect text with regard to illegal content, such as discrimination. The approach described in the paper extends the existing state-of-the-art in the field and uses a multi-task learning architecture based on deep learning (DL). The result is relevant for every social media platform that wants to improve compliance by proactively detecting problematic comments and statements.

The remainder of the paper is structured as follows: Section 2 describes related work and similar approaches in the domain of sentiment analysis to detect illegal statements. It also introduces some recent work on utilizing DL in this matter. Section 3 introduces the multi-label system by describing the dataset that is used, the algorithms, and the topology of the architecture. The experimental setup is described in Section 4, while the results are discussed in Section 5. Based on the results, the limitations and potential arising research directions are described in Section 6.

\section{Related Work}
The computer-assisted analysis of legal documents is highly relevant and has attracted researchers for quite some time. Sentiment analysis, also called opinion mining, is an important field of research, not only in the legal domain \cite{conrad2007opinion}. Pang et al. \cite{pang2008opinion} provide a comprehensive overview of different approaches to sentiment analysis across various domains. This paper deals with sentiment analysis in comments from various online sources in order to detect illegal comments. Hutto \& Gilbert \cite{gilbert2014vader} came up with a rule based system, called VADER, to analyze social media text. A deep dive on supervised sentiment analysis based on various multilingual web texts was made by Boiy \& Moens \cite{boiy2009machine}. They used hand-crafted features to train three different ML classifiers (support vector machines (SVM), multinomial naive bayes, and maximum entropy classifier) on the given task. Even though there are several existing works on this topic, among others \cite{hong2016malicious, Paltoglou:2012:TMD:2337542.2337551, Paltoglou2014, sood2012automatic}, hardly any attempt has been made in the legal domain by utilizing DL. 

\section{Illegal Content Detection System}

In this section, we are covering the datasets which we used, and a detail description of the Multi-Task Multi-Embedding algorithm that we have used.

\subsection{Datasets}

\begin{figure}[htb!]
\centering
\includegraphics[width=\linewidth]{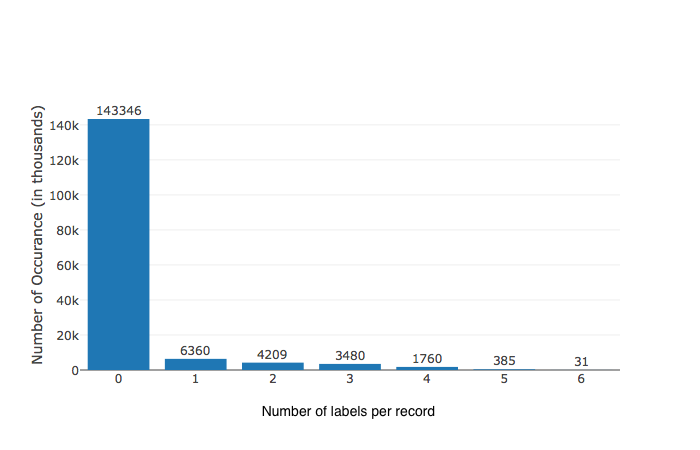}
\caption{Number of multi-label occurrence on the training dataset}
\label{fig:multiLabelOccurrence}
\end{figure}

\textbf{Multi-Label Illegal Comments:} The main dataset was provided by Jigsaw which is a team within Google through one of Kaggle competitions \cite{JigsawDataset2018}. Each sample has a text comment which is labeled with one or more labels among six labels: Toxic, Severe Toxic, Obscene, Threat, Insult and Identity Hate.  The dataset is divided into a training set and a test set. The number of samples of the training dataset and test dataset is 159571 and 63978 respectively. Figure~\ref{fig:multiLabelOccurrence} shows the number of multilabel occurrences on the training dataset.

\begin{figure}[htb!]
\centering
\includegraphics[width=\linewidth]{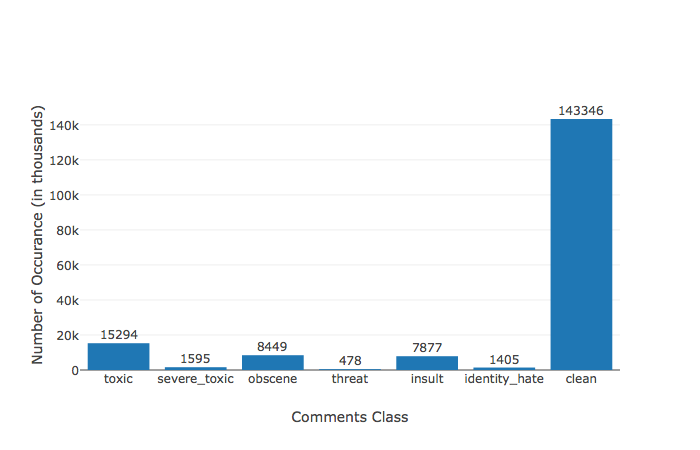}
\caption{Number of comment occurrence on the training dataset for each class }
\label{fig:classOccurrence}
\end{figure}

The biggest challenge of this dataset is imbalance class distribution, where the number of true classes is between 0.3\% and 10\%. In these cases most machine learning algorithms tend to give very high accuracy because it is easier for the algorithms to just produce one value either 1 or 0 to minimize the loss function. However, in this case, it crucially important to predict the toxicity of the comments and not just predict every comment as clean comment. This is a major problem in the legal domain where the number of positive or negative samples for many problems are very low. In Figure~\ref{fig:classOccurrence}, the number of comments occurrence of each class on the training dataset is presented. For example, the number of threat comments is about 0.3\% which make it difficult for the algorithm to predict it correctly.

The dataset contains text from various languages. The top 30 unicode scripts occurrence on the training dataset is presented in Table~\ref{table:languageOccurrence}. This is another problem since most of the modern machine learning algorithm uses specific language word representations like word embedding. Which means the model will just set all the other languages words to zero. However, since the number of Non-English comments was less than 0.5\% and as shown on the word cloud almost all of the major illegal words were in English, we only focused on English comments.

\begin{table}[htb!]
\centering
\caption{Top 30 unicode scripts occurrence on the train dataset}
\label{table:languageOccurrence}
\begin{tabular}{|l|l|l|l|l|l|}
\hline
\textbf{Language} & \textbf{Occurrence} & \textbf{Language} & \textbf{Occurrence} & \textbf{Language} & \textbf{Occurrence} \\ \hline
Latin             & 159564              & Bengali           & 25                  & Kannada           & 4                   \\ \hline
Greek             & 456                 & Runic             & 22                  & Lao               & 4                   \\ \hline
Han               & 344                 & Hangul            & 21                  & Gujarati          & 3                   \\ \hline
Cyrillic          & 272                 & Ethiopic          & 20                  & Telugu            & 3                   \\ \hline
Arabic            & 78                  & Georgian          & 17                  & Tibetan           & 3                   \\ \hline
Hiragana          & 66                  & Tamil             & 9                   & Oriya             & 2                   \\ \hline
Katakana          & 55                  & Thai              & 8                   & Sinhala           & 2                   \\ \hline
Devanagari        & 52                  & Gurmukhi          & 7                   & Bopomofo          & 1                   \\ \hline
Inherited         & 50                  & Armenian          & 6                   & Malayalam         & 1                   \\ \hline
Hebrew            & 48                  & Khmer             & 5                   & Syriac            & 1                   \\ \hline
\end{tabular}
\end{table}

\begin{figure}[htb!]
\begin{subfigure}{0.5\textwidth}
  \centering
  \includegraphics[width=\linewidth]{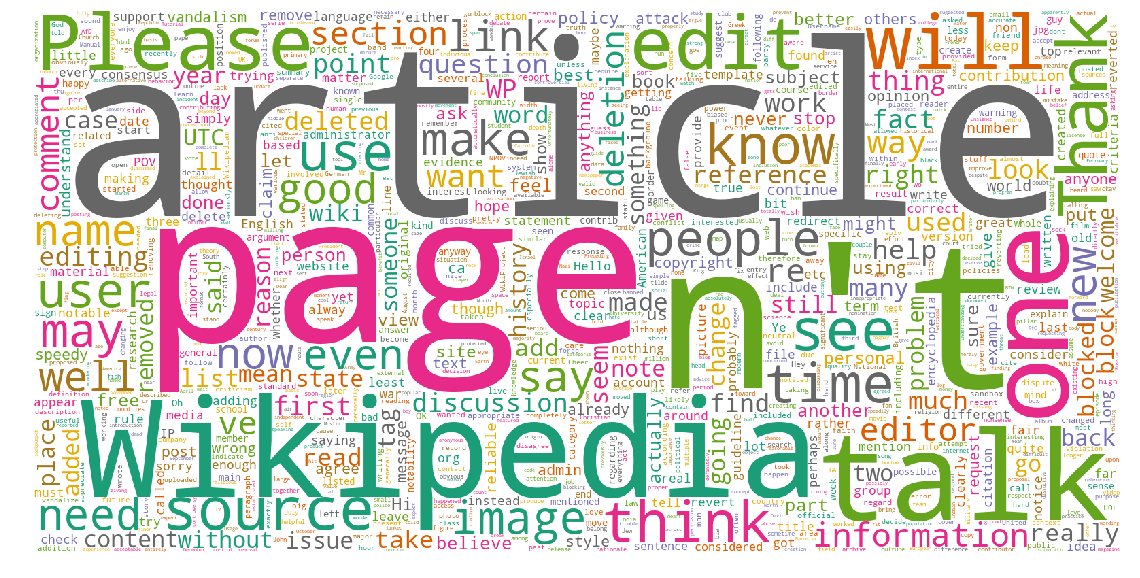}
  \caption{Clean}
  \label{fig:wordCloudClean}
\end{subfigure}%
\begin{subfigure}{.5\textwidth}
  \centering
  \includegraphics[width=\linewidth]{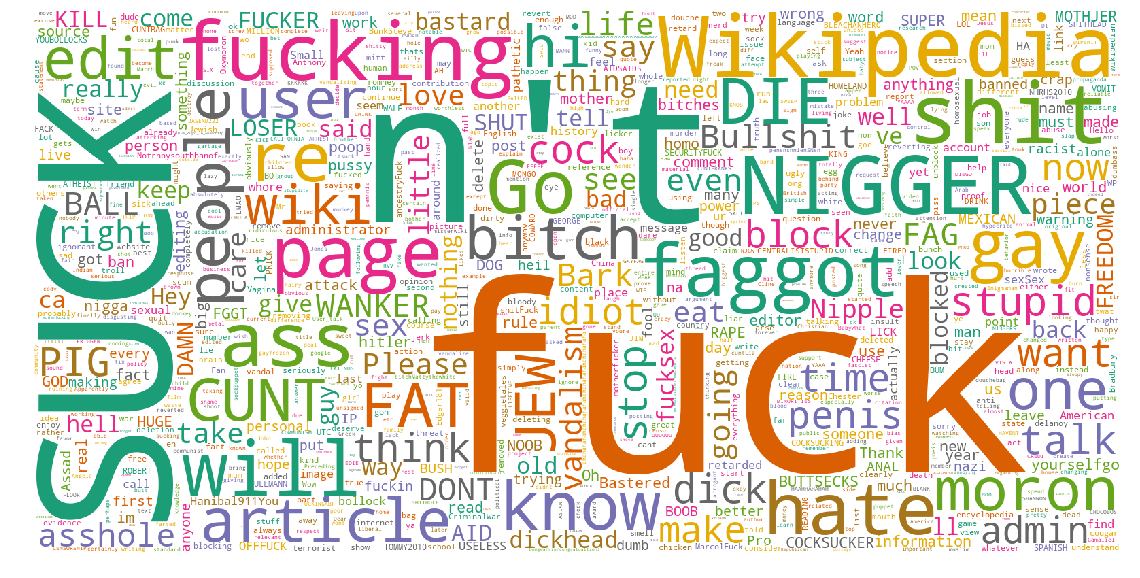}
  \caption{Toxic}
  \label{fig:wordCloudToxic}
\end{subfigure}
\begin{subfigure}{.5\textwidth}
  \centering
  \includegraphics[width=\linewidth]{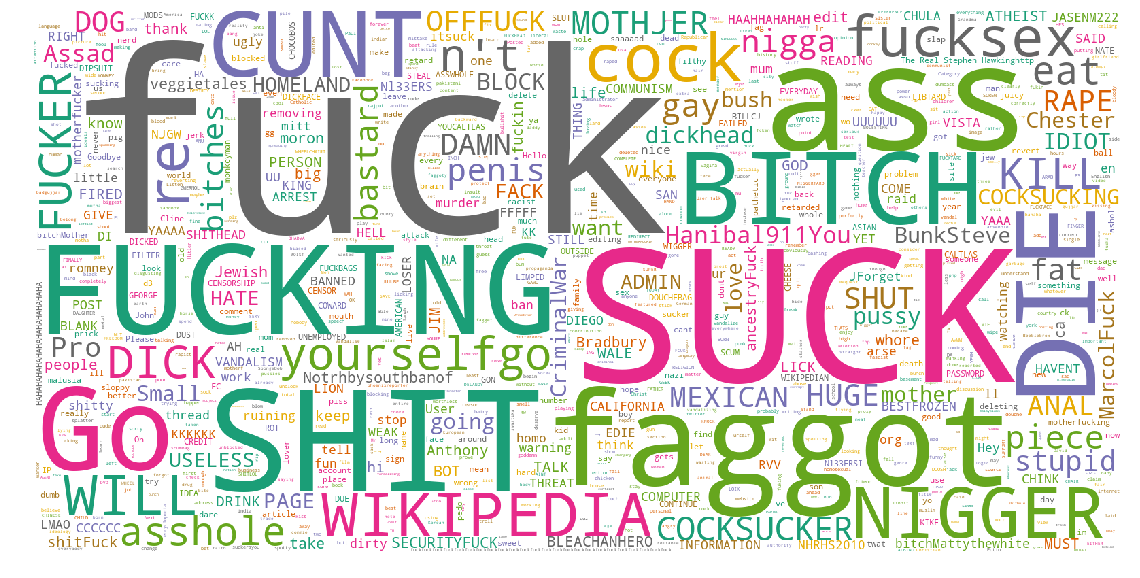}
  \caption{Severe Toxic}
  \label{fig:wordCloudSevereToxic}
\end{subfigure}%
\begin{subfigure}{.5\textwidth}
  \centering
  \includegraphics[width=\linewidth]{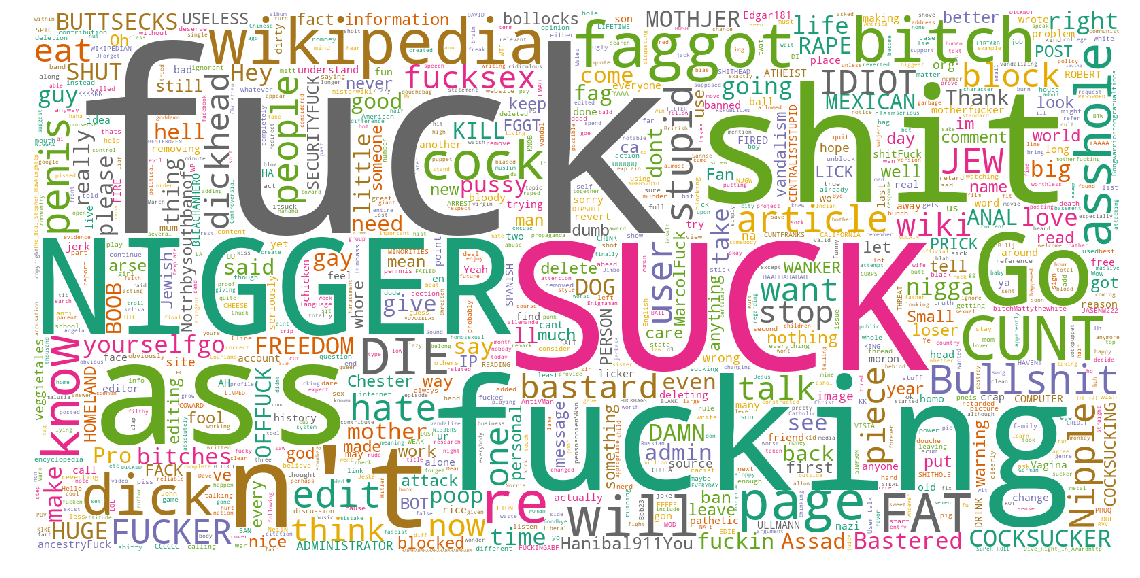}
  \caption{Obsence}
  \label{fig:wordCloudObsence}
\end{subfigure}
\begin{subfigure}{.5\textwidth}
  \centering
  \includegraphics[width=\linewidth]{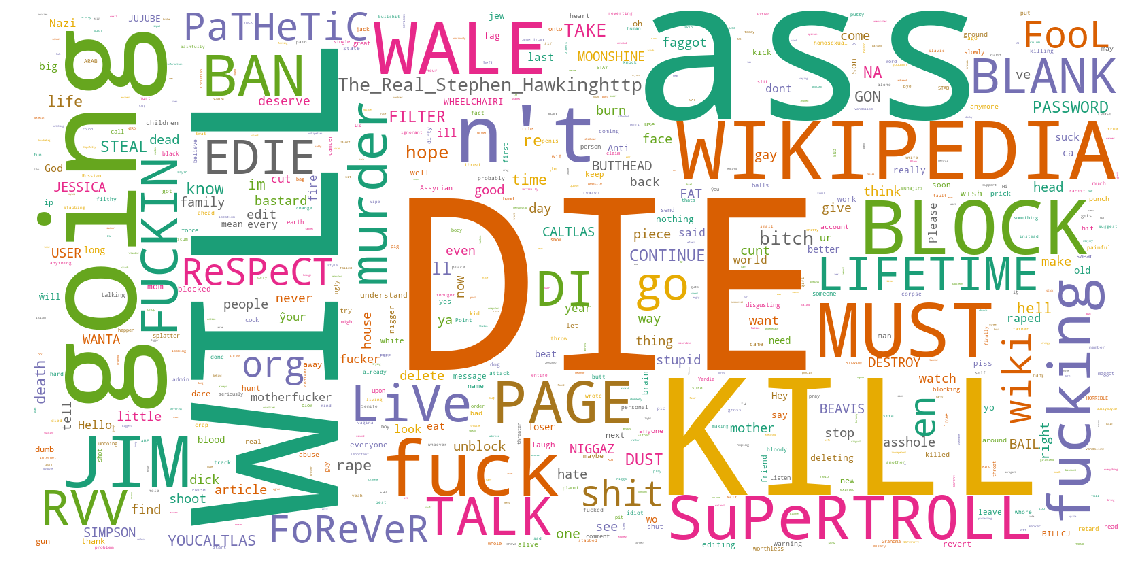}
  \caption{Threat}
  \label{fig:wordCloudThreat}
\end{subfigure}%
\begin{subfigure}{.5\textwidth}
  \centering
  \includegraphics[width=\linewidth]{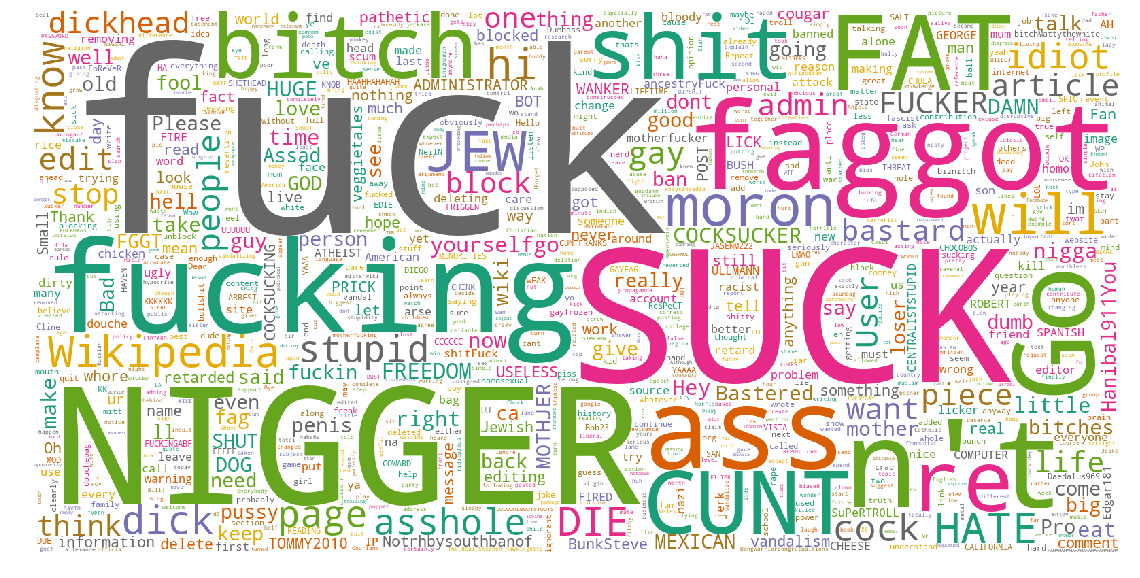}
  \caption{Insult}
  \label{fig:wordCloudInsult}
\end{subfigure}
\begin{subfigure}{.5\textwidth}
  \centering
  \includegraphics[width=\linewidth]{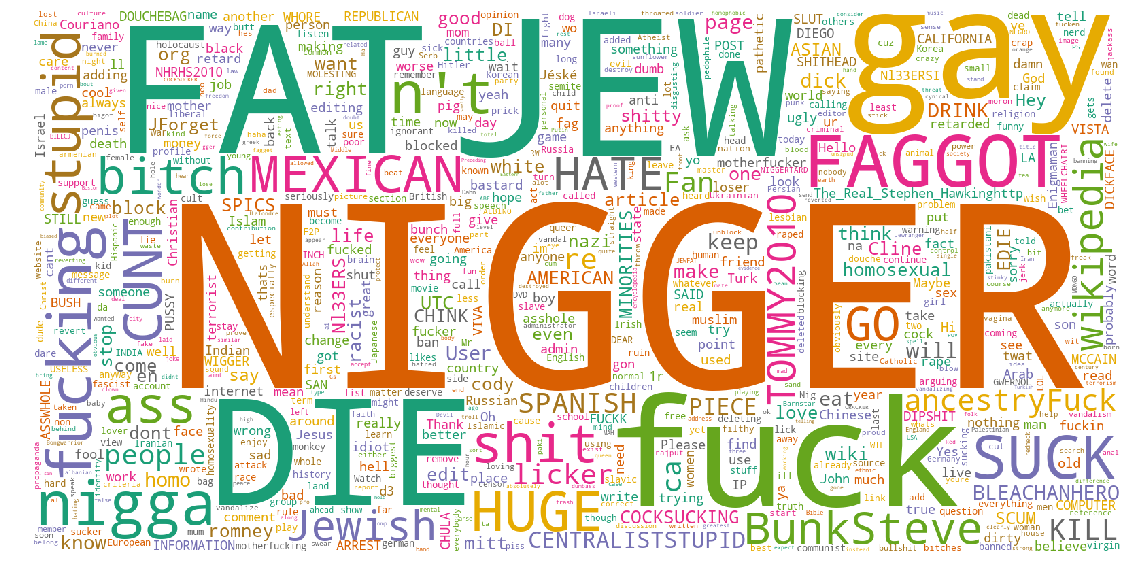}
  \caption{Identity Hate}
  \label{fig:wordCloudIdentityHate}
\end{subfigure}%
\begin{subfigure}{.5\textwidth}
  \centering
  \includegraphics[width=\linewidth]{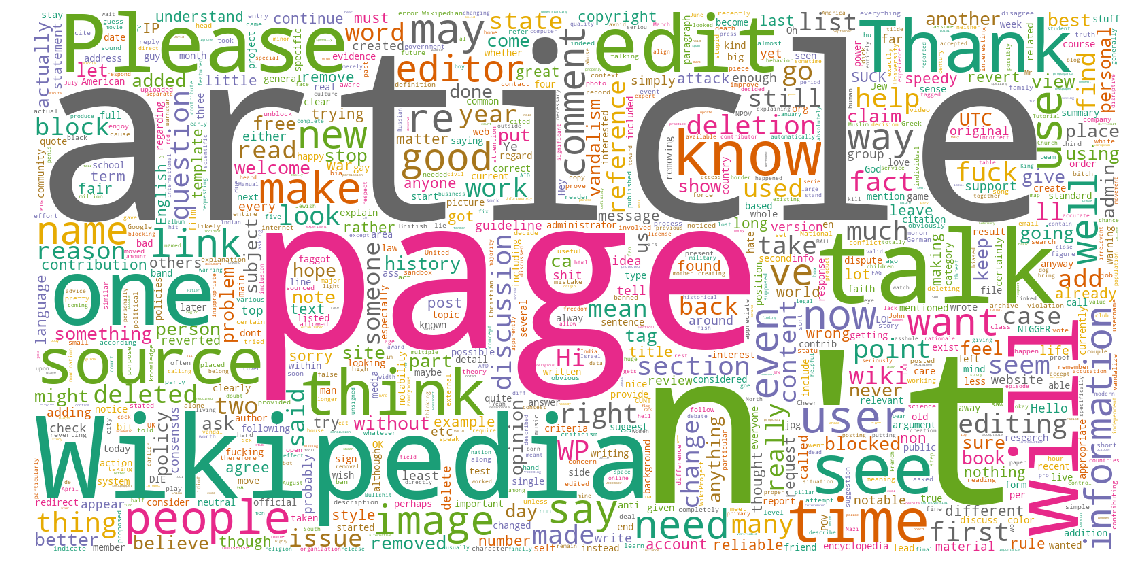}
  \caption{All Combined}
  \label{fig:wordCloudCombined}
\end{subfigure}
\caption{Single gram word cloud of the top 1000 word occurrence for each comments class}
\label{fig:wordCloud}
\end{figure}

A word cloud for each class is presented in Figure~\ref{fig:wordCloud} for the top 1000 word occurrence. Since the illegal comments were collected from wiki page comments, we can notice that the top keywords on clean comments are "Wikipedia, page and article". A second example, on "Identity Hate" class, the top keywords target specific race or specific religion like "Nigger or Jew". A third example, on "Threat" class, the top keywords were life threatening words like "Die or Kill". In Figure \ref{fig:wordCloudCombined}, we can see a word cloud of over all top 1000 word occurrence, and clearly none of the most occurred words on the six labels appear. This makes the machine learning algorithms difficult to predict them.

\hfill\break
\textbf{Toxic, Attack and Aggression Comments:} Additionally, three datasets were used to analyze if it is beneficial for the main dataset to join training it with these three datasets using Multi-task training. These specific datasets were chosen because generally in multi-task training, the closer the tasks the better the results. All of them were also provided by Jigsaw team. The number of samples of the toxic, attack and aggression datasets are 159686, 115864, 115864 samples. These three datasets also suffer from the imbalance class distribution problem. 

\subsection{Multi-Task Multi-Embedding Model}
\begin{figure}[htb!]
\centering
\includegraphics[width=\linewidth]{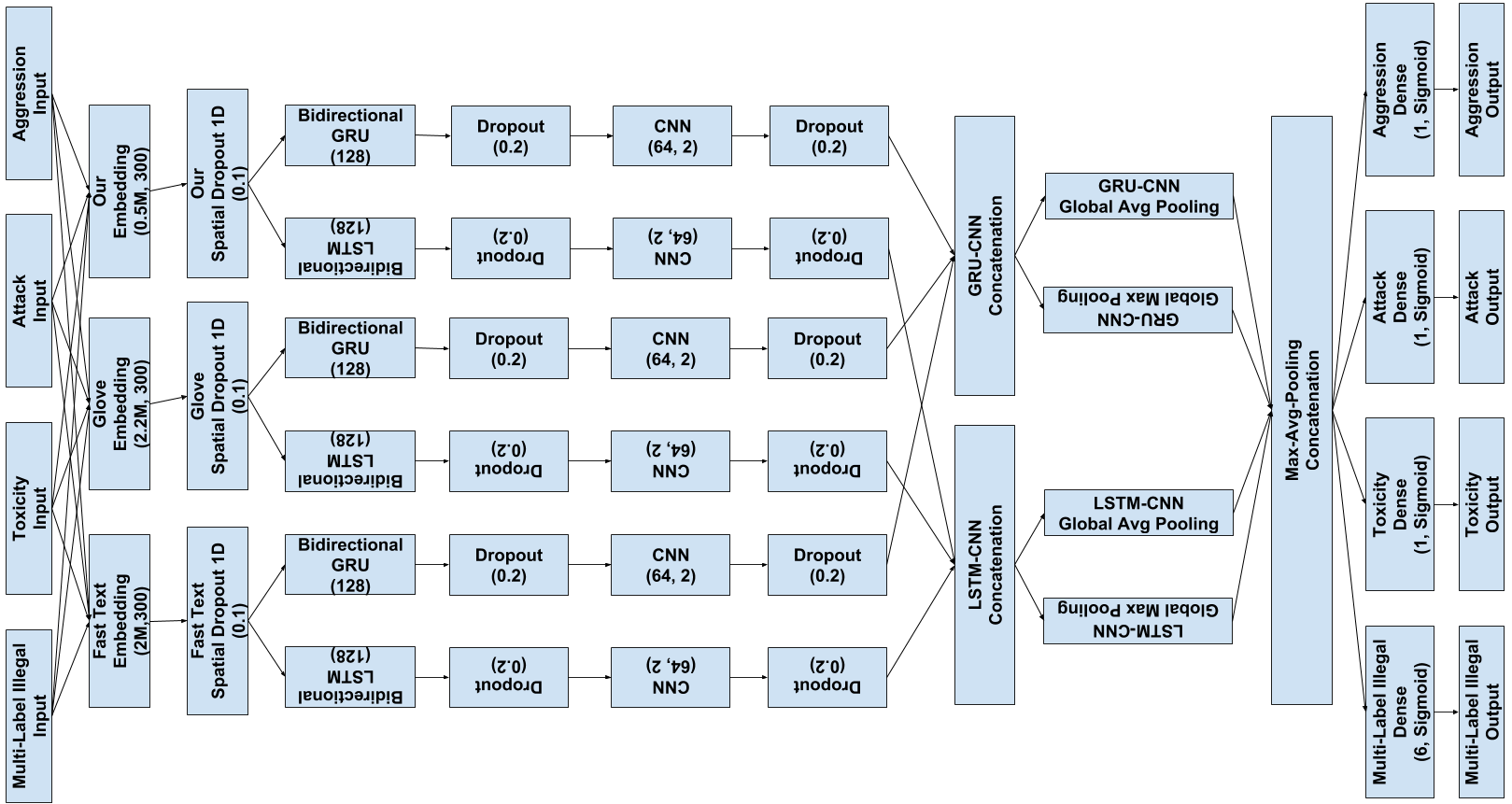}
\caption{Multi-Task Multi-Embedding Deep Learning Architecture}
\label{fig:modelArchitecture}
\end{figure}

Figure~\ref{fig:modelArchitecture} shows the Model Architecture, where the model receives four different inputs from four different datasets (Main Multi-Label Illegal, Toxic, Attack and Aggression) and produce four different outputs for each task, while the whole layers are hard shared among all tasks~\cite{ruder2017overview}. Simply, hard sharing means all the internal learning parameters are shared among all tasks. The model can be divided into the following layers:

\subsubsection{Word Representations Layer}
Each input goes through three different word embedding layers (FastTexts~\cite{joulin2016bag} , Glove~\cite{pennington2014glove} and our Glove). The Fast Text is 2 million word vectors trained on Common Crawl with dimension 300, while Glove is 2.2 million word vectors trained on Common Crawl with dimension 300. Furthermore, we trained a custom Glove model with comments from English Wikipedia talk pages~\cite{wulczyn2017ex} which produced 0.5 million word vectors with dimension 300.
Each word embedding model is followed by 20\% Spatial 1D Dropout, which simply tries to hide some words representations' during training to reduce over fitting.
Two ideas are behind using different word representation, first is to capture different semantic and syntactic features of words, second is to cover as many words as possible by training them on different unstructured text sources.   

\subsubsection{Recurrent Neural Network (RNN) Layers}
Each word embedding layer is followed by two parallel Bidirectional RNN layers. The first is a GRU and second one is LSTM layer. Both layers has 128 neurons and is followed by 20\% dropout. This creates a total of 3 GRU layers and 3 LSTM layers. The main benefit of using these RNN layers is to allow the model to capture the relationship between the sequence of words. 

\subsubsection{Convolutional Neural Network (CNN) Layers}
Afterwards, a separate CNN layer takes the output of each separate RNN layer, which creates a total of 6 CNN layers. The hyper parameters of each CNN are as follows: the number of filters is 64, the kernel size is 2, and the activation function was RELU. Each one is followed by 20\% dropout. 
Afterward The output of the CNN layers is concatenated into 2 outputs. The first and second concatenation concatenates the output of the 3 CNN layers coming after the LSTM layers and the output of the 3 CNN layers coming after the GRU layers respectively.

\subsubsection{Pooling Layers}
The output of each concatenation of the CNN layers goes through separate Max Pooling and Average Pooling. Afterwards, the output of the 2 Max Pooling and the 2 Average Pooling layers are concatenated together.

\subsubsection{Dense Layers}
Finally, separate 4 dense layers take the output of the concatenated pooling layers to produce the final outputs. The first dense layer produces 6 outputs for each label of the main illegal comment dataset (Toxic, Severe Toxic, Obscene, Threat, Insult and Identity Hate), while the second, third and fourth dense layers produce only 1 output for each class on each complementary datasets (Toxic, Attack and Aggression).

\section{Experimental Settings}

\subsection{Training Details}
Several single task algorithms were tested against the multi-task multi-embedding algorthim. All the single task algorithms were trained only on the main dataset, while the multi-task multi-embedding algorithm were trained on at least two data sets together.
Some standard classic approach algorithms were tested like Linear Regression (LR) and Decision Trees (DT). Other single task deep learning algorithms which were tested were reported by top participants in the Kaggle competition including: BI-GRU, CNN and BI-RNN-CNN models. The BI-GRU models is 2 layers of bi-directional GRU with 128 units. The CNN models consist of 4 parallel CNN layers "300 neurons and  kernel sizes are 2,3,4 and 5 respectively" followed by global max pooling, then concatenated together and followed by a final dense layer with 36 neurons. The BI-RNN-CNN models consist of 2 parallel GRU and LSTM "128 neurons" layers, followed separate CNN layers "64 neurons", followed by max and average and max pooling and then concatenated together. 

Generally, all the models were trained until they converged using early stop to prevent over-fitting. For neural network based models a mini-batch stochastic gradient decent was used, with a binary cross-entropy loss function. Furthermore, the Adam optimizer was used to optimize the loss function on all of the neural network models with learning rate $ 1e^{-3} $. L1-norm or L2-norm regularization were not used, but dropout was used as discussed on the previous section.
For our Multi-Task model at each training step we trained the model for the same batch size of each problem sequentially.

All Models were trained using one computer with one Nvidia Titan XP GPU with 12 GB Ram, 20 CPU cores and 128 Memory Ram.

\subsection{Metrics}
As stated in the previous subsection the main dataset is skewed, which means the accuracy can't be a reliable measurement. That is why we choose the precision, recall and F1 score to give us a more reliable measurement. The equations for precision, recall and F1 score is shown in equation ~\ref{equ:precisionEqu}, ~\ref{equ:RecallEqu} and ~\ref{equ:F1Equ} respectively.

\begin{equation}
Precision = \frac{\#True Positives}{\#True Positives + \#False Positives}
\label{equ:precisionEqu}
\end{equation}
\begin{equation}
Recall = \frac{\#True Positives}{\#True Positives + \#False Negatives}
\label{equ:RecallEqu}
\end{equation}
\begin{equation}
F_{1} = \frac{Precision \times Recall}{Precision + Recall}
\label{equ:F1Equ}
\end{equation}

\section{Results and Discussion}

Table~\ref{table:MLResults} shows the result of the tested algorithms regarding precision, recall and F1 score matrices. All of the algorithms were able to reach an average 99\% F1 score for all the negative labels "Clean Comment" however, the problems arise with the positive labels "illegal comments" where the average F1 score ranges between 38\% and 59\%. 
The Linear regression algorithm using TF-IDF performed the worst with an average positive F1 score 38\%. However, the decision tree using TF-IDF, which is another classical ML algorithm performed well compared to the deep learning models with F1 score 54\%, despite the fact it doesn't use any word embedding layer which transfer a lot of information for all the tokens, after being trained in big unstructured text datasets. The reason is that Decision Trees (DT) algorithms don't affect too much with imbalanced class data compared to other machine learning approaches. Both the BI-GRU and CNN models using fastText performed lower than the DT with an average F1 score 53\% and 52\%. The best model among all the single task approaches was the BI-RNN-CNN model using fastText with an average F1 score 58\%. However, by changing the embedding layer with Glove and our pre-trained model the BI-RNN-CNN F1 score decreased with average F1 score 54\% and 52\%.

The multi-task multi-embedding algorithm when used by combining the main data set with either toxic, attack or aggression didn't out-perform the best single task model with average F1 score 55\%, 56\% and 56\%. However, when the 4 data sets were combined it out-performed the best single task model with an average F1 score 59\%.

First observation, the balance of samples per label affect dramatically the performance of all algorithms. All models were able to perfectly predict a clean comment which was not the case with illegal comments, because the number of clean samples is almost about 97\% of the data set.

Second observation, multi-task algorithms with more data sets generalize much better than when only few data sets are chosen. Despite the fact that the model couldn't outperform all the F1 scores of every label, but it did generalize better on the overall average F1 score.

Third observation, multi-task algorithms which were fed with the 4 datasets had much higher true positive Rate (Recall), especially on the labels with very few samples (severe toxic, threat and identity hate). Which in this case makes it a better choice since the algorithm tends to be sensitive to classify the illegal comments.



\begin{table}[htp!]
\centering
\caption{Precision, Recall and F1 score of base line machine learning models versus our proposed Multi-Task Multi-Embedding Model}
\label{table:MLResults}
\resizebox{\columnwidth}{!}{%

\begin{tabular}{|l|c|c|c|c|c|c|c|c|c|c|c|c|c|}
\hline
\multicolumn{3}{|l|}{\textbf{Archticture}}                                                   & \multicolumn{7}{c|}{\textbf{Single Task}}                                                                  & \multicolumn{4}{c|}{\textbf{Multi-Task}}             \\ \hline
\multicolumn{3}{|l|}{\textbf{Model}}                                                         & \textbf{LR} & \textbf{DT} & \textbf{BI-GRU} & \textbf{CNN} & \multicolumn{3}{c|}{\textbf{BI-RNN-CNN}}      & \multicolumn{4}{c|}{\textbf{Multi-Emb RNN-CNN}}             \\ \hline
\multirow{4}{*}{\textbf{Datasets}}       & \multicolumn{2}{l|}{\textbf{Multi-Label Illegal}} & X           & X           & X               & X            & X             & X             & X             & X             & X    & X             & X             \\ \cline{2-14} 
                                         & \multicolumn{2}{l|}{\textbf{Toxic}}               &             &             &                 &              &               &               &               & X             &      &               & X             \\ \cline{2-14} 
                                         & \multicolumn{2}{l|}{\textbf{Attack}}              &             &             &                 &              &               &               &               &               & X    &               & X             \\ \cline{2-14} 
                                         & \multicolumn{2}{l|}{\textbf{Agression}}           &             &             &                 &              &               &               &               &               &      & X             & X             \\ \hline
\multirow{4}{*}{\textbf{Word Embedding}} & \multicolumn{2}{l|}{\textbf{TF-IDF}}               & X           & X           &                 &              &               &               &               &               &      &               &               \\ \cline{2-14} 
                                         & \multicolumn{2}{l|}{\textbf{FastText}}            &             &             & X               & X            & X             &               &               &X               & X    & X             & X             \\ \cline{2-14} 
                                         & \multicolumn{2}{l|}{\textbf{Glove}}               &             &             &                 &              &               & X             &               &X               & X    & X             & X             \\ \cline{2-14} 
                                         & \multicolumn{2}{l|}{\textbf{Our}}                 &             &             &                 &              &               &               & X             &X               & X    & X             & X             \\ \hline
\multirow{6}{*}{\textbf{Toxic}}          & \multirow{2}{*}{\textbf{P}}      & \textbf{0}     & 0,95        & 0,98        & 0,98            & 0,99         & 0,98          & 0,98          & 0,98          & 0,98          & 0,98 & 0,98          & 0,98          \\ \cline{3-14} 
                                         &                                  & \textbf{1}     & 0,78        & 0,63        & 0,6             & 0,55         & 0,64          & 0,62          & 0,64          & 0,61          & 0,6  & 0,6           & 0,64          \\ \cline{2-14} 
                                         & \multirow{2}{*}{\textbf{R}}      & \textbf{0}     & 0,98        & 0,95        & 0,94            & 0,93         & 0,95          & 0,95          & 0,95          & 0,94          & 0,94 & 0,94          & 0,95          \\ \cline{3-14} 
                                         &                                  & \textbf{1}     & 0,56        & 0,77        & 0,85            & 0,87         & 0,81          & 0,81          & 0,78          & 0,83          & 0,82 & 0,83          & 0,8           \\ \cline{2-14} 
                                         & \multirow{2}{*}{\textbf{F1}}     & \textbf{0}     & 0,97        & 0,96        & 0,96            & 0,95         & 0,97          & 0,96          & 0,97          & 0,96          & 0,96 & 0,96          & 0,96          \\ \cline{3-14} 
                                         &                                  & \textbf{1}     & 0,65        & 0,7         & 0,7             & 0,67         & \textbf{0,71} & 0,7           & 0,7           & \textbf{0,71} & 0,7  & 0,7           & \textbf{0,71} \\ \hline
\multirow{6}{*}{\textbf{Severe toxic}}   & \multirow{2}{*}{\textbf{P}}      & \textbf{0}     & 1           & 1           & 1               & 1            & 1             & 1             & 1             & 1             & 1    & 1             & 1             \\ \cline{3-14} 
                                         &                                  & \textbf{1}     & 0,37        & 0,35        & 0,45            & 0,39         & 0,51          & 0,43          & 0,45          & 0,45          & 0,45 & 0,35          & 0,35          \\ \cline{2-14} 
                                         & \multirow{2}{*}{\textbf{R}}      & \textbf{0}     & 1           & 1           & 1               & 1            & 1             & 1             & 1             & 1             & 1    & 1             & 1             \\ \cline{3-14} 
                                         &                                  & \textbf{1}     & 0,2         & 0,25        & 0,27            & 0,31         & 0,21          & 0,36          & 0,34          & 0,2           & 0,22 & 0,39          & 0,41          \\ \cline{2-14} 
                                         & \multirow{2}{*}{\textbf{F1}}     & \textbf{0}     & 1           & 1           & 1               & 1            & 1             & 1             & 1             & 1             & 1    & 1             & 1             \\ \cline{3-14} 
                                         &                                  & \textbf{1}     & 0,26        & 0,29        & 0,34            & 0,35         & 0,3           & \textbf{0,39} & \textbf{0,39} & 0,28          & 0,29 & 0,37          & 0,38          \\ \hline
\multirow{6}{*}{\textbf{Obscene}}        & \multirow{2}{*}{\textbf{P}}      & \textbf{0}     & 0,97        & 0,98        & 0,98            & 0,99         & 0,99          & 0,98          & 0,98          & 0,98          & 0,98 & 0,98          & 0,98          \\ \cline{3-14} 
                                         &                                  & \textbf{1}     & 0,88        & 0,68        & 0,66            & 0,59         & 0,64          & 0,73          & 0,73          & 0,69          & 0,68 & 0,69          & 0,7           \\ \cline{2-14} 
                                         & \multirow{2}{*}{\textbf{R}}      & \textbf{0}     & 1           & 0,98        & 0,98            & 0,97         & 0,97          & 0,98          & 0,98          & 0,98          & 0,98 & 0,98          & 0,98          \\ \cline{3-14} 
                                         &                                  & \textbf{1}     & 0,5         & 0,72        & 0,75            & 0,8          & 0,78          & 0,68          & 0,68          & 0,73          & 0,73 & 0,73          & 0,71          \\ \cline{2-14} 
                                         & \multirow{2}{*}{\textbf{F1}}     & \textbf{0}     & 0,98        & 0,98        & 0,98            & 0,98         & 0,98          & 0,98          & 0,98          & 0,98          & 0,98 & 0,98          & 0,98          \\ \cline{3-14} 
                                         &                                  & \textbf{1}     & 0,64        & 0,7         & 0,7             & 0,68         & 0,7           & \textbf{0,71} & 0,7           & \textbf{0,71} & 0,7  & \textbf{0,71} & \textbf{0,71} \\ \hline
\multirow{6}{*}{\textbf{Threat}}         & \multirow{2}{*}{\textbf{P}}      & \textbf{0}     & 1           & 1           & 1               & 1            & 1             & 1             & 1             & 1             & 1    & 1             & 1             \\ \cline{3-14} 
                                         &                                  & \textbf{1}     & 1           & 0,65        & 0,49            & 0,31         & 0,58          & 0,44          & 0,51          & 0,38          & 0,35 & 0,36          & 0,4           \\ \cline{2-14} 
                                         & \multirow{2}{*}{\textbf{R}}      & \textbf{0}     & 1           & 1           & 1               & 1            & 1             & 1             & 1             & 1             & 1    & 1             & 1             \\ \cline{3-14} 
                                         &                                  & \textbf{1}     & 0           & 0,28        & 0,12            & 0,14         & 0,47          & 0,28          & 0,13          & 0,51          & 0,59 & 0,52          & 0,6           \\ \cline{2-14} 
                                         & \multirow{2}{*}{\textbf{F1}}     & \textbf{0}     & 1           & 1           & 1               & 1            & 1             & 1             & 1             & 1             & 1    & 1             & 1             \\ \cline{3-14} 
                                         &                                  & \textbf{1}     & 0,01        & 0,39        & 0,2             & 0,2          & \textbf{0,52} & 0,34          & 0,2           & 0,43          & 0,44 & 0,42          & 0,48          \\ \hline
\multirow{6}{*}{\textbf{Insult}}         & \multirow{2}{*}{\textbf{P}}      & \textbf{0}     & 0,97        & 0,98        & 0,98            & 0,98         & 0,98          & 0,98          & 0,98          & 0,98          & 0,97 & 0,97          & 0,98          \\ \cline{3-14} 
                                         &                                  & \textbf{1}     & 0,83        & 0,74        & 0,72            & 0,6          & 0,73          & 0,68          & 0,69          & 0,76          & 0,8  & 0,76          & 0,77          \\ \cline{2-14} 
                                         & \multirow{2}{*}{\textbf{R}}      & \textbf{0}     & 1           & 0,99        & 0,99            & 0,97         & 0,99          & 0,98          & 0,98          & 0,99          & 0,99 & 0,99          & 0,99          \\ \cline{3-14} 
                                         &                                  & \textbf{1}     & 0,4         & 0,58        & 0,65            & 0,73         & 0,64          & 0,68          & 0,65          & 0,57          & 0,54 & 0,54          & 0,56          \\ \cline{2-14} 
                                         & \multirow{2}{*}{\textbf{F1}}     & \textbf{0}     & 0,98        & 0,98        & 0,98            & 0,98         & 0,98          & 0,98          & 0,98          & 0,98          & 0,98 & 0,98          & 0,98          \\ \cline{3-14} 
                                         &                                  & \textbf{1}     & 0,54        & 0,65        & \textbf{0,68}   & 0,66         & \textbf{0,68} & \textbf{0,68} & 0,67          & 0,65          & 0,64 & 0,63          & 0,65          \\ \hline
\multirow{6}{*}{\textbf{Identity hate}}  & \multirow{2}{*}{\textbf{P}}      & \textbf{0}     & 0,99        & 0,99        & 0,99            & 0,99         & 1             & 0,99          & 0,99          & 0,99          & 0,99 & 0,99          & 1             \\ \cline{3-14} 
                                         &                                  & \textbf{1}     & 0,73        & 0,78        & 0,72            & 0,59         & 0,59          & 0,73          & 0,8           & 0,61          & 0,6  & 0,54          & 0,54          \\ \cline{2-14} 
                                         & \multirow{2}{*}{\textbf{R}}      & \textbf{0}     & 1           & 1           & 1               & 1            & 1             & 1             & 1             & 1             & 1    & 1             & 0,99          \\ \cline{3-14} 
                                         &                                  & \textbf{1}     & 0,09        & 0,39        & 0,48            & 0,51         & 0,58          & 0,32          & 0,3           & 0,47          & 0,55 & 0,48          & 0,64          \\ \cline{2-14} 
                                         & \multirow{2}{*}{\textbf{F1}}     & \textbf{0}     & 0,99        & 1           & 1               & 1            & 1             & 1             & 1             & 1             & 1    & 0,99          & 0,99          \\ \cline{3-14} 
                                         &                                  & \textbf{1}     & 0,16        & 0,52        & 0,57            & 0,54         & \textbf{0,59} & 0,44          & 0,43          & 0,53          & 0,57 & 0,51          & 0,58          \\ \hline
\multirow{2}{*}{\textbf{Total Average}}          & \multirow{2}{*}{\textbf{F1}}     & \textbf{0}     & 0,99        & 0,99        & 0,99            & 0,99         & 0,99          & 0,99          & 0,99          & 0,99          & 0,99 & 0,99          & 0,99          \\ \cline{3-14} 
                                         &                                  & \textbf{1}     & 0,38        & 0,54        & 0,53            & 0,52         & 0,58          & 0,54          & 0,52          & 0,55          & 0,56 & 0,56          & \textbf{0,59} \\ \hline
\end{tabular}

}

\end{table}


\section{Conclusion}

Multi-task deep learning models are a promising new technology that could pave the way for more deep learning applications in the legal domain by transferring knowledge from large datasets to small datasets. We investigate the transfer learning capabilities of a particular multi-task architecture on a classification task on the publicly available toxic-comments Kaggle challenge dataset. 
We explore the toxic-comments dataset in depth and identify the imbalanced classes as a major challenge.Our multi-task approach does not significantly improve upon single-task models in terms of F1-score. However, it improves the recall score significantly for labels with a low number of samples, which is extremely important to this use-case. A key result is that we can observe a significant improvement in F1-score when adding additional, related datasets to train the multi-task model.

These promising results encourage us to further investigate the transfer learning capabilities of multi-task models. In particular, it is unclear, what deep neural network architectures are most suitable for applications in the legal domain. It will be necessary to explore different combinations of tasks, datasets, architectures and hyper parameter selections to better understand the transfer learning capabilities of transfer learning especially with multi-task models.

\section{Acknowledgements}
We gratefully acknowledge the support of NVIDIA Corporation with the donation of the Titan XP Pascal GPU used for this research.


\begin{bibliography}{bibliography/references.bib}
\bibliographystyle{plain}
\end{bibliography}

\end{document}